\documentclass[aps,prb,twocolumn,superscriptaddress,showpacs]{revtex4}
\usepackage{graphicx}
\usepackage{latexsym}
\usepackage{amssymb}
\usepackage{amsmath}
\usepackage{amsfonts}
\usepackage{bm}
\usepackage{multirow}
\usepackage{color}
\usepackage{comment}

\newcommand{\beq}{\begin{equation}}
\newcommand{\eeq}{\end{equation}}
\newcommand{\beqn}{\begin{eqnarray}}
\newcommand{\eeqn}{\end{eqnarray}}

\DeclareMathAlphabet{\mathbbold}{U}{bbold}{m}{n}

\makeatletter
\newcommand\xleftrightarrow[2][]{%
\ext@arrow 9999{\longleftrightarrowfill@}{#1}{#2}}
\newcommand\longleftrightarrowfill@{%
\arrowfill@\leftarrow\relbar\rightarrow} \makeatother

\def\sfT{\mathsf{T}}
\def\cl{\mathcal{C}\ell}
\def\NLSM{{\rm NL}\sigma{\rm M}}

\begin{document}

\title{Interacting Topological Insulators with Synthetic Dimensions}

\author{Chao-Ming Jian}
\affiliation{ Station Q, Microsoft Research, Santa Barbara,
California 93106-6105, USA} \affiliation{Kavli Institute of
Theoretical Physics, Santa Barbara, CA 93106, USA}

\author{Cenke Xu}
\affiliation{Department of Physics, University of California,
Santa Barbara, CA 93106, USA}

\date{\today}
\begin{abstract}

Recent developments of experimental techniques have given us
unprecedented opportunities of studying topological insulators in
high dimensions, while some of the dimensions are ``synthetic", in
the sense that the effective lattice momenta along these synthetic
dimensions are controllable periodic tuning parameters. In this
work, we study interaction effects on topological insulators with
synthetic dimensions. We show that although the free fermion band
structure of high dimensional topological insulators can be
precisely simulated with the ``synthetic techniques", the generic
interactions in these effective synthetic topological insulators
are qualitatively different from the local interactions in
ordinary condensed matter systems. And we show that these special
but generic interactions have unexpected effects on topological
insulators, namely they would change (or reduce) the
classification of topological insulators {\it differently} from
the previously extensively studied local interactions.

\end{abstract}

\maketitle

\section{Introduction}


Ever since the proposal of the analogue of quantum Hall effects in
four spatial dimensions~\cite{4dqh}, the topological states of
matters in higher dimensions have attracted a great deal of
theoretical interests. The classification of free fermion
topological insulators (TI) and superconductors (TSC) in all
dimensions (the so-called ``10-fold way") was a milestone in our
understanding of non-interacting fermionic
systems~\cite{ludwigclass1,ludwigclass2,kitaevclass}. Later, a
great progress in understanding strongly interacting bosonic
states of matter was achieved through the classification and
description of bosonic symmetry protected topological states,
which can also be generalized to higher
dimensions~\cite{wenspt,wenspt2,kapustin1,kapustin2,freed1,freed2,wenso,bixu,youxu,youxu2,jianye,wenSM,xuGUT,xu16}.

Until recently, the study of higher dimensional topological
insulators (TI) and its bosonic analogues was of pure theoretical
interests only. However, the study of higher dimensional TIs has
gained important experimental relevance recently. For example, the
four dimensional quantum Hall insulator (or the four dimensional
Chern insulator) was successfully constructed
experimentally~\cite{4dqh2,4dqhex} with two out of its four
spatial dimensions ``synthetic". In fact, ``synthetic" dimensions
are experimentally realized as periodic tuning parameters that can
be identified as the corresponding lattice momenta in these
dimensions. In general, what such synthetic-dimension techniques
directly realize is a Hamiltonian of the form $ \hat{H}(\vec{p}) =
\sum_{<i,j>} \hat{H}_{i,j}(\vec{p}) c^\dagger_{i}c_{j}$, where
$\vec{p}$ is a $\delta = (D-d)$-component tuning parameters and
$i,j$ label the sites on the $d$-dimensional physical (optical)
lattice. Once we identify $\vec{p}$ as the lattice momenta in the
synthetic dimensions, the entire system can be viewed as a
$D$-dimensional non-interacting fermion system. Such synthetic
construction in principle can give us experimental realization of
all classes of non-interacting fermion TI in any dimension. The
same perspective, $i.e.$ viewing lattice momenta in certain
dimensions as tuning parameters, has been used by theorists to
connects TIs in different dimensions~\cite{qiTI}, and also to
construct topological
semimetals~\cite{weylashvin,weylleon,weyldai,weylreview,weylex1,weylex2,diracex1}.

Interestingly, interactions can drastically change the
classification of TIs and TSCs with certain symmetries. It was
first discovered in Ref.~\onlinecite{fidkowski1,fidkowski2} that
some TIs nontrivial in the non-interacting limit can be
trivialized by interactions, which exemplified the importance of
interaction in altering or more precisely ``reducing" the
classification of TIs. Subsequent studies have shown the same
effect of interaction, i.e. the reduction of classification, in
TIs and TSCs in all
dimensions~\cite{qiz8,yaoz8,zhangz8,chenhe3B,senthilhe3,youxu2,xuGUT,xu16,xufb,fuz4,hermelez4,z42,wenz4}.
These conclusions were made under the assumptions of {\it
spatially local} interactions that preserve the crucial symmetries
which define the TIs.

In contrast, the generic interaction on a synthetic TI is
fundamentally different from that on the ordinary TIs given the
``synthetic" nature of the extra dimensions. The interaction must
be local in the $d$-dimensional real space lattice, and also local
in the $\delta = (D-d)$-dimensional synthetic {\it momentum}
space, $i.e.$ under adiabatic tuning, at each value of the tuning
parameter $\vec{p}$ the system has an interaction in the
$d$-dimensional real space labelled by $\vec{p}$:
$H_{int}(\vec{p})$. This allows us to use the ``dimensional
reduction" procedure of Ref.~\onlinecite{qiTI} even with
interactions, which is a method that we will exploit in this work.

\begin{figure}[tbp]
\begin{center}
\includegraphics[width=220pt]{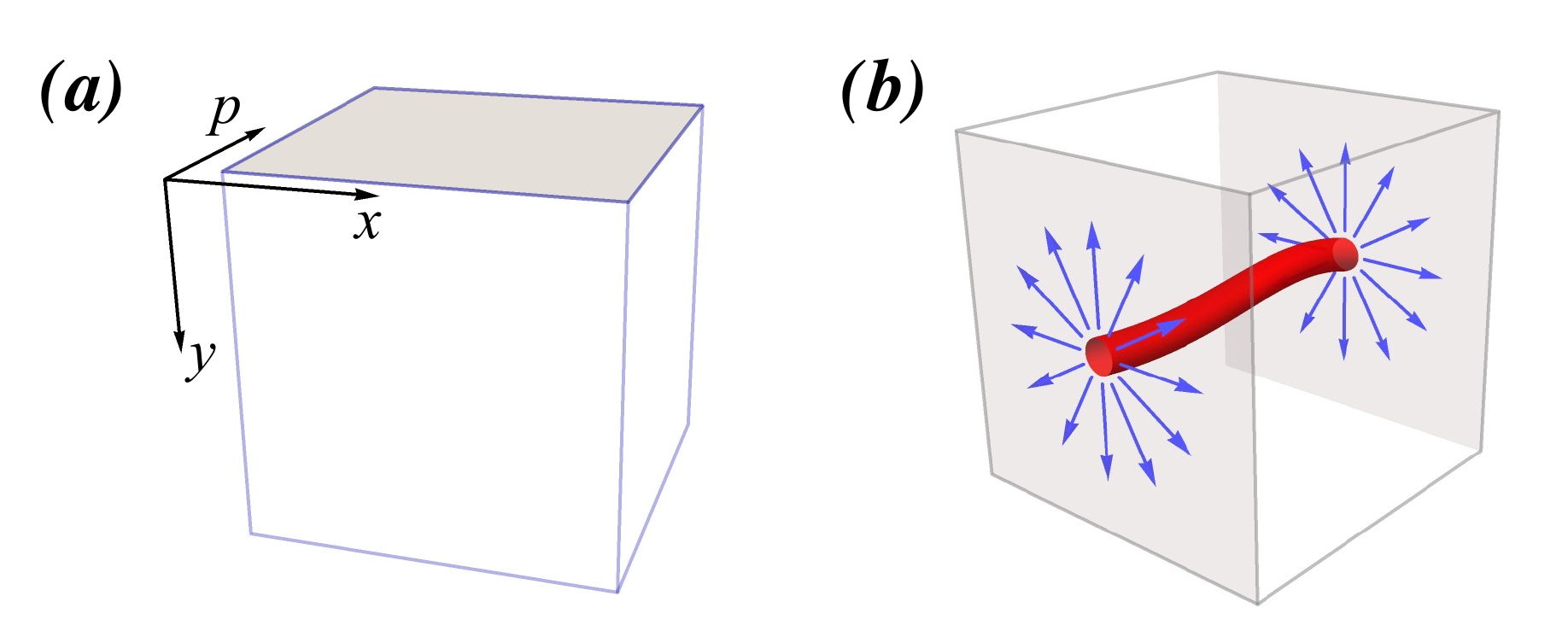}
\caption{($a$) The $D=3$ synthetic TI with $d=2$, the lattice
momentum $p$ is physically a tuning parameter. We always consider
the boundary parallel with the synthetic dimensions. Here, the
exposed boundary is in the $(x,p)$ plane with fixed $y$
coordinate. ($b$) The vortex line of the complex bosonic order
parameter $\phi$ in Eq.~\ref{phi}, and there is one nonchiral $1d$
Majorana fermion mode localized in the core of the vortex line
(Eq.~\ref{H1d}). } \label{fig}
\end{center}
\end{figure}

A generic interaction $H_{int}(\vec{p})$ in a synthetic TI is no
longer completely local in the effective $D$-dimensional real
space: it is local in the physical dimensions, but nonlocal in the
synthetic spatial dimensions. It appears that TIs with a nonlocal
interaction is not even definable, since a nonlocal interaction
would easily mix the edge states at two opposite boundaries of the
TIs. But in the current situation, the interaction is only
nonlocal along the synthetic directions in the effective
$D$-dimensional space. And as long as we are considering
boundaries parallel to the $\delta=(D-d)$-dimensional synthetic
space, namely the $\delta$-component synthetic momenta are still
conserved at the boundary, the two opposite boundaries will not be
mixed by the nonlocal interaction (Fig.~\ref{fig}a). This is the
most natural choice of boundary in the effective $D$-dimensional
synthetic TI, as we simply need to choose the physical boundary of
the system while keeping the tuning parameter $\vec{p}$ periodic.
Then one can still study the fate of the edge states at one single
boundary, and the effective $D$-dimensional system can be called a
nontrivial TI as long as this boundary remains gapless under the
interaction.

\begin{center}
\begin{tabular}{ |c|c|c|c|c|c|}
\hline $(D, \delta)$     & $(1, 0)$     & $(1, 1)$      &$(3, 0)$
& $(3, 1)$     & $(3, 2)$     \\ \hline
Classification, $U(1) \times Z_2^T$ &$\mathbb{Z}_4$&$\mathbb{Z}_2$ &$\mathbb{Z}_8$&$\mathbb{Z}_4$&$\mathbb{Z}_2$\\
\hline
\end{tabular}
\label{table1}
\end{center}

\begin{center}
\begin{tabular}{ |c|c|c|c|c|c| }
\hline $(D, \delta)$   &$(2, 0)$      & $(2, 1)$      &$(4, 0)$ &
$(4, 1)$     & $(4, 2)$     \\ \hline
Classification, $U(1) \times Z_2$ &$\mathbb{Z}_4$&$\mathbb{Z}_2$ &$\mathbb{Z}_8$&$\mathbb{Z}_4$&$\mathbb{Z}_2$\\
\hline
\end{tabular}
\label{table2}
\end{center}

Our main results are the following: (1) The interaction reduced
classification for TIs with symmetry $U(1) \times Z_2^T$ at total
odd dimensions $D=2n+1$ and synthetic dimensions $\delta$ is given by
$\mathbb{Z}_{2^{n+2-\delta}}$; (2) The interaction reduced
classification for {\it non-chiral} TIs with symmetry $U(1) \times
Z_2$ at total even dimensions $D=2n$ and synthetic dimensions $\delta$
is given by $\mathbb{Z}_{2^{n+1-\delta}}$. We select these
symmetries because TIs defined with these symmetries, as we will
discuss in the next few sections, will be strongly affected by the
interaction. In the tables above, we list the results for
$D=2,3,4$ explicitly, which will be discussed in detail in Sec.
\ref{sec:D=2} \ref{sec:D=3} and \ref{sec:D=4}. In Sec.
\ref{sec:D=2n} and \ref{sec:D=2n+1}, we will discuss the
classification reduction of interacting synthetic non-chiral TIs
with $U(1) \times Z_2$ symmetry in general even dimensions $D=2n$
and that of interacting synthetic TIs with $U(1) \times Z^T_2$
symmetry in general odd dimensions $D=2n+1$.


\section{Synthetic TI with $D=2$}
\label{sec:D=2}

We will start with the example of $D = 2$ synthetic TI, with $d =
\delta = 1$, namely one of the two dimensions is synthetic. The
synthetic non-interacting Chern insulator with $D=2$ and $\delta=1$
has been studied previously~\cite{synqh}. One particular type of
two dimensional TIs that we know will be strongly influenced by
interaction and reduce its classification is the {\it non-chiral}
TI with $U(1)\times Z_2$ symmetry~\cite{fuz4,hermelez4}, where
physically the $Z_2$ is usually the reflection symmetry about the
$z$ axis, which becomes an on-site symmetry in the two dimensional
plane. The minimal version of this TI is basically two copies of
Chern insulators with opposite Chern numbers, and also opposite
eigenvalues ($\pm 1$) under the $Z_2$ symmetry operation. Its $1d$
edge state has the Hamiltonian \beqn H = \int dx \ v \left(
\psi^\dagger_1 i
\partial_x \psi_1 - \psi^\dagger_2 i
\partial_x \psi_2 \right). \eeqn The charge conservation $U(1)$ symmetry
and the $Z_2$ symmetry act on the boundary fermions as \beqn U(1):
(\psi_1, \psi_2) \rightarrow e^{i\theta} (\psi_1,\psi_2), \cr \cr
Z_2: (\psi_1, \psi_2) \rightarrow (\psi_1, - \psi_2). \eeqn One
can see that for arbitrary copies of the edge states, any fermion
bilinear mass term at the edge will mix left and right moving
fermions, and hence break the $Z_2$ symmetry, while any
fermion-pairing mass gap would break the $U(1)$ symmetry. Hence
without interaction the classification of this TI is $\mathbb{Z}$.
It was shown that under local interaction the classification of
this $U(1)\times Z_2$ TI is reduced to
$\mathbb{Z}_4$~\cite{fuz4,hermelez4}. This conclusion has two
related implications:

(1). Four copies of the one dimensional edge states can be gapped
out by local interactions without breaking either $U(1)$ or $Z_2$
symmetry, which was directly shown in Ref.~\onlinecite{fuz4};

(2). For four copies of the system, the TI-to-trivial phase
transition in the two dimensional bulk can be avoided under
interaction, namely the non-interacting TI phase and the trivial
phase can be adiabatically connected with interaction while
keeping the bulk gap open. This can be understood using the
Chalker-Coddington network picture of the bulk topological
transition~\cite{cc,cc2}: the transition between two phases with
the same symmetry and no bulk ground state degeneracy can be
interpreted as the percolation of their interface (domain wall).
If the interface is fully gapped, then these two phases can be
adiabatically connected without closing the bulk gap.

Let us now investigate what if the dimension along the boundary is
synthetic. In this case, the boundary Hamiltonian including the
interaction will be \beqn H = \sum_p H_p, \ \ H_p = v p
\left(\psi_{1,p}^\dagger \psi_{1,p} - \psi^\dagger_{2,p}\psi_{2,p}
\right) + H_{int}(p). \label{edge1dp}\eeqn Because $p$ is the
synthetic momentum, it is actually a tuning parameter, $H_{int}(p)$ is local in the synthetic momentum space, $i.e.$ for
each value of $p$ there is an interaction labelled by $p$. Then
for each $p$, solving $H_p$ is equivalent to solving a zero
dimensional problem with finite dimensional Hilbert space. The
original gapless point $p=0$ becomes a level crossing of two
states: one state has $N_{1,p} = \psi_{1,p}^\dagger \psi_{1,p} =
1$, $N_{2,p} = 0$ and the other state has $N_{1,p} = 0$, $N_{2,p}
= 1$.

\begin{figure}[tbp]
\begin{center}
\includegraphics[width=200pt]{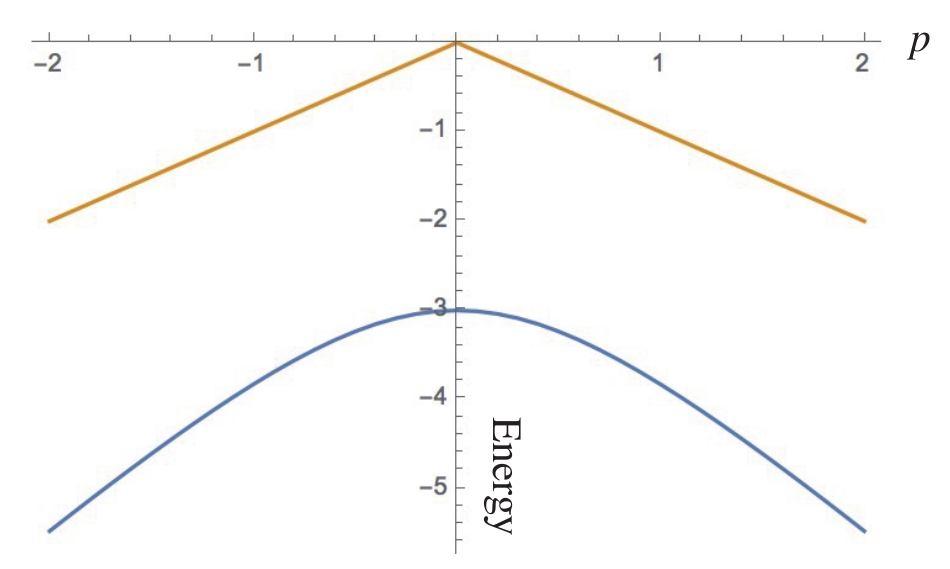}
\caption{The exact many-body spectrum (including the ground state
and the first excited state) of the Hamiltonian Eq.~\ref{edge1dp2}
with the interaction chosen to be Eq.~\ref{int}. The ground state
is non-degenerate and is separated from the excited states by a
finite energy gap for all values of $p$.} \label{fig: SynTi2D}
\end{center}
\end{figure}

If $H_{int}(p)$ preserves the $U(1)\times Z_2$ symmetry, then for
a single copy of Eq.~\ref{edge1dp}, there is no way to avoid the
level crossing, because $H_{int}(p)$ can always be recombined into
a function of $N_{1,p}$ and $N_{2,p}$, and changing the filling of
$N_{1,p}$ and $N_{2,p}$ will lead to level crossing. But for two
copies of the Eq.~\ref{edge1dp}, the story is different. Let us
perform a basis transformation of the second copy of the system so that the edge state Hamiltonian reads \beqn H_p = v p \left(
\psi_{1,p}^\dagger \tau^z \psi_{1,p} - \psi^\dagger_{2,p} \tau^z
\psi_{2,p} \right) + H_{int}(p). \label{edge1dp2}\eeqn Our goal is
to show that for certain choice of $H_{int}(p)$, the ground state
in the limit $p > 0$ and $ p < 0$ of Eq.~\ref{edge1dp2} can be
connected adiabatically without closing the ground state gap. The
following choice of $H_{int}(p)$ will suffice: \beqn
H_{int}(p) = J \vec{S}_{p,+} \cdot \vec{S}_{p,-}, \ \
\vec{S}_{p,\pm} = \psi^\dagger_{p,\pm} \vec{\sigma} \psi_{p,\pm},
\label{int}\eeqn where $\vec{\sigma}$ are three Pauli matrices
that act on the fermion index $(1,2)$, and $\psi_{p,\pm}$ are
fermion modes with eigenvalues $\pm 1$ of $\tau^z$. The direct
computation of the spectrum of Eq.~\ref{edge1dp2} plotted in Fig.
\ref{fig: SynTi2D} confirms that the interaction $H_{int}$ can
indeed ensure a finite gap for all values of $p$ for the
Hamiltonian Eq. \ref{edge1dp2}.

This analysis implies that the two dimensional non-chiral TI with
$U(1) \times Z_2$ symmetry has its classification reduced from
$\mathbb{Z}$ to $\mathbb{Z}_2$ under interaction if one of the
spatial dimensions is synthetic, which is different from the
$\mathbb{Z}$ to $\mathbb{Z}_4$ reduction as was discussed
previously.

\section{Synthetic TI with $D=3$}
\label{sec:D=3}

Now let us consider the case with $D = 3$, and $\delta=1$, namely
we are considering an effective $D=3$ dimensional TI with one
synthetic dimension. One type of three dimensional TI whose
classification is changed by interaction is the TI defined by
symmetry $U(1) \times Z_2^T$. One tight-binding model of this TI
is $\hat{H} = \sum_{\vec{k}} c^\dagger_{\vec{k}} H(\vec{k})
c_{\vec{k}}$, and $H(\vec{k})$ is \beqn H(\vec{k}) = - t
\left(\sum_{i = 1}^3 \Gamma^i \sin k_i - \Gamma^4(h - \sum_{i=1}^3
\cos k_i) \right). \eeqn Here, we use the following convention of the Gamma
matrices $\Gamma^1 = \sigma^{30}$, $\Gamma^2 = \sigma^{10}$,
$\Gamma^3 = \sigma^{22}$, $\Gamma^4 = \sigma^{21}$, $\Gamma^5 =
\sigma^{23}$, and $\sigma^{ab} = \sigma^a \otimes \sigma^b$, with
$\sigma^0 = \mathbf{1}_{2\times 2}$. The anti-unitary
time-reversal $Z_2^T$ symmetry acts on the fermion operators as
$\mathcal{T}: c_k \rightarrow i \Gamma^5 c_{k}^\dagger$, combined
with a complex conjugation. This model is essentially two copies
of the topological superconductors $^3$He-B phase. Note that
whether $\mathcal{T}^2$ is $+1$ or $-1$ no longer matters in this
case as the sign of $\mathcal{T}^2$ can be changed by the $U(1)$
rotation. In the literatures this $U(1)$ symmetry is often
referred to as the spin $U(1)$ symmetry since it commutes with the
time-reversal (for example, see Ref.~\onlinecite{wenz4}).

At the (for example) XY boundary, the system has a $2d$ gapless
Dirac fermion with Hamiltonian \beqn H = \int d^2x \ v
\psi^\dagger(-i \sigma^3 \partial_x - i \sigma^1 \partial_y )\psi,
\label{edge2dx} \eeqn The time-reversal symmetry acts on the
boundary two-component Dirac fermion as \beqn \mathcal{T}: \psi
\rightarrow i\sigma^y \psi^\dagger. \eeqn It is well-known that
without any interaction, the classification of this TI is
$\mathbb{Z}$~\cite{ludwigclass1,ludwigclass2,kitaevclass}, while
under ordinary local interactions, the classification of this TI
is reduced to $\mathbb{Z}_8$~\cite{senthilhe3}, $i.e.$ 8 copies of
this TI will be rendered trivial under local interaction, or
equivalently the edge state can be gapped by local interaction
without developing nonzero expectation value of any fermion
bilinear operator, which is forbidden by the $U(1) \times Z_2^T$.

Now let us investigate what happens if the $y-$direction is a
synthetic dimension, and consider multiple copies of the boundary
Hamiltonian. Then the boundary Hamiltonian at each synthetic
momentum $p$ is \beqn H_p = \int dx \ \sum_{a = 1}^N v
\psi^\dagger_{a, p}(-i \sigma^3
\partial_x + \sigma^1 p ) \psi_{a, p} + H_{int}(p), \label{edge2d}
\eeqn where $H_{int}(p)$ is a $U(1)\times Z_2^T$ symmetry allowed
flavor mixing interaction term that is parameterized by $p$.
Notice that the time-reversal symmetry does not mix fermion
operators labelled by different momenta, $i.e.$ time-reversal does
not mix systems with different parameter $p$. The entire
Hamiltonian Eq.~\ref{edge2d} can be viewed as $N-$copies of $1d$
{\it interacting} topological insulator with the same $U(1)\times
Z_2^T$ symmetry (which again has $\mathbb{Z}$ classification in
the non-interacting limit), tuned close to its transition to the
trivial insulator, which corresponds to $N$ copies of $1d$ Dirac
fermions with Dirac mass $p$. The synthetic momentum $p$ is
precisely the Dirac mass that tunes the system across the
topological-trivial transition. Let us emphasize again that this
does not apply to the ordinary interacting TI if all dimensions
are physical dimensions, because different momenta will be mixed
by the ordinary local interactions.

Now the problem readily reduces to the interaction effects on the
$1d$ TI with $U(1)\times Z_2^T$ symmetry, which has been studied
and well understood. It was shown that the $1d$ TI with
$U(1)\times Z_2^T$ symmetry, though have $\mathbb{Z}$
classification without interaction, is reduced to $\mathbb{Z}_4$
classification under local interaction~\cite{wenz4,z42}. This
$\mathbb{Z}$ to $\mathbb{Z}_4$ reduction under interaction is also
consistent with the classification of bosonic symmetry protected
topological phases: two copies of the TIs under interaction can be
adiabatically connected to the Haldane phase by gapping out the
single particle excitations~\cite{xufb}, and it is well-known that
two copies of coupled Haldane phases become a trivial
phase~\cite{wenspt,wenspt2}.

The observation above implies that when $N=4$ in Eq.~\ref{edge2d},
the phase transition between the two limits $p > 0$ and $p < 0$
can be avoided by turning on interaction, $i.e.$ the two regions
in the phase diagrams can be adiabatically connected under
interaction without closing the gap, and the original critical
point $p = 0$ is rendered gapped and nondegnerate by interaction.
This observation leads to the conclusion that the classification
of the $D=3$ TI with $U(1)\times Z_2^T$ symmetry is reduced to
$\mathbb{Z}_4$ instead of $\mathbb{Z}_8$ if one of the three
spatial dimensions is synthetic.

\begin{figure}[tbp]
\begin{center}
\includegraphics[width=240pt]{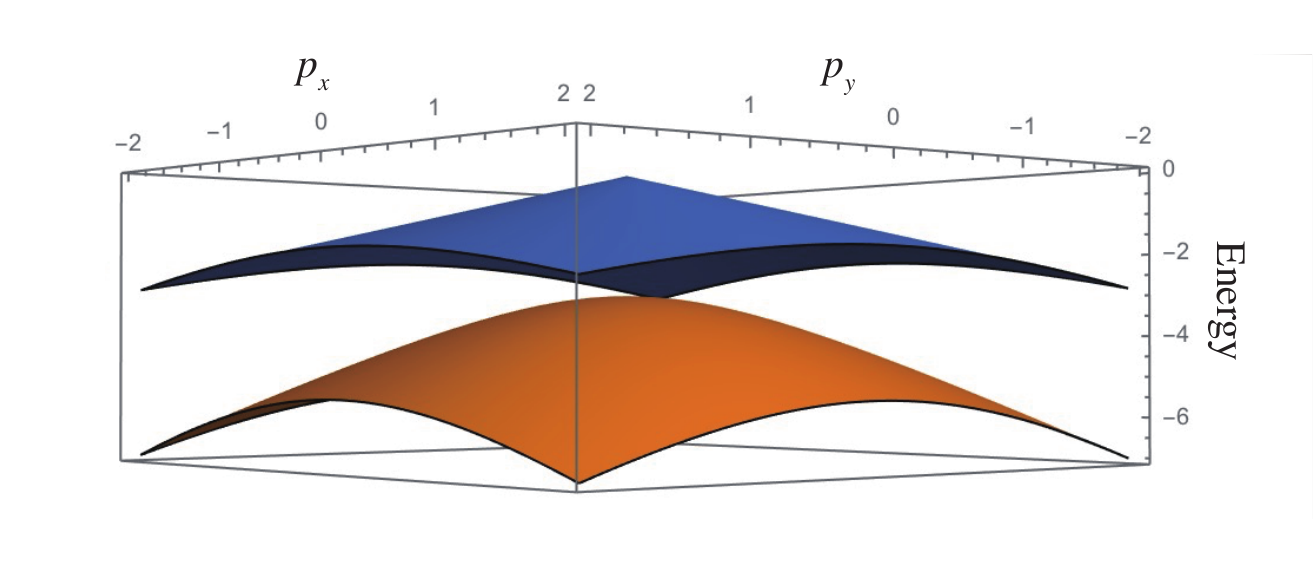}
\caption{ The exact many-body spectrum (including the ground state
and the first excited state) of the Hamiltonian
$\psi^\dagger_{\vec{p}} \left( \sigma^3 p_x + \sigma^1 p_y \right)
\otimes\tau^z \psi_{\vec{p}} + \vec{S}_{p,+} \cdot \vec{S}_{p,-}
$. The ground state is non-degenerate and is separated from the
excited states by a finite energy gap for all values of $p_x$ and
$p_y$.
 } \label{fig: SynTi3D}
\end{center}
\end{figure}

If we choose $D=3$, $\delta = 2$, namely two out of the three
directions are synthetic dimensions, and we consider a two
dimensional boundary whose both directions are synthetic (let us
label them as the $x$ and $y$ directions), then at each $p_x$ and
$p_y$ the edge state is a two component complex fermion
$\psi_{\vec{p}}$. Again the problem at the boundary reduces to
solving a zero dimensional system with two tuning parameter $p_x$,
$p_y$. For $N = 2$ copies of the edge states, at every synthetic
momentum $\vec{p}$, the free fermion part of the boundary
Hamiltonian reads \beqn H_{\vec{p}} = v \psi^\dagger_{\vec{p}}
\left( \sigma^3 p_x + \sigma^1 p_y \right) \otimes\tau^z
\psi_{\vec{p}}, \label{edge2p}\eeqn where again we have performed a basis
transformation for $\tau^z = -1$ component of the edge state. Then
the same interaction as Eq.~\ref{int}, $H_{int}(p) = J
\vec{S}_{p,+} \cdot \vec{S}_{p,-}$ can gap out the entire boundary
without leading to any ground state degeneracy (see Fig. \ref{fig: SynTi3D}). Hence when there
are two synthetic dimensions, the $D=3$ TI with $U(1)\times Z_2^T$
is reduced to a $\mathbb{Z}_2$ classification.

To summarize this section, for a $D = 3$ interacting synthetic TI
with $U(1)\times Z_2^T$ symmetry and different choices of $\delta
= D - d$, its classification is \beqn \mathbb{Z}_8, \ (\delta =
0); \ \  \ \mathbb{Z}_4, \ (\delta = 1); \ \ \ \mathbb{Z}_2, \
(\delta = 2). \eeqn

One can also discuss the simplest example of $D = 1$ TI with
$U(1)\times Z_2^T$ symmetry. If the only dimension is synthetic,
namely the lattice momentum along this dimension is actually a
tuning parameter, although it is unnatural to discuss edge state
of the synthetic dimension, the TI can still be defined as whether
there must be a gap closing transition between the TI and the
trivial insulator or not. In the non-interacting limit, near the
critical point between the TI and trivial insulator, the bulk
Hamiltonian takes exactly the same form as Eq.~\ref{edge1dp}, and
the time-reversal symmetry acts as $\mathcal{T}: \psi_p
\rightarrow i \sigma^y \psi_p^\dagger$. The mass term that tunes
the topological-to-trivial transition is $m \psi^\dagger_p
\sigma^x \psi_p$. Then one can show that with two copies of the
system, an interaction $H_{int}(p)$ similar to Eq.~\ref{int} would
adiabatically connect the original topologically nontrivial TI to
the trivial phase. Thus interaction reduces the classification of
the synthetic $D=1$ TI with $U(1)\times Z_2^T$ symmetry to
$\mathbb{Z}_2$.

\section{Synthetic TI with $D=4$}
\label{sec:D=4}

Now we discuss the higher dimensional (total dimension $D=4$) TI
that cannot be realized in lab without using the ``synthetic"
techniques. First of all, the classification of the $D=4$ quantum
Hall insulator itself is not reduced by interaction, due to the
chiral anomaly at its three dimensional boundary, and the anomaly
matching condition~\cite{anomalymatching,anomalymatching2}. Thus
we will consider the $D=4$ non-chiral TI with a $U(1)\times Z_2$
symmetry. Like its $D=2$ analogue, this TI is simply two copies of
the original $D=4$ quantum Hall state with opposite Chern numbers,
and they have eigenvalues $+1$ and $-1$ respectively under the
$Z_2$ symmetry operation.

Since this $D=4$ TI was not discussed much in the past, we will
first consider the case where all the dimensions are physical,
$i.e.$ $D = d=4$. The $3d$ boundary of this TI is two copies of
Weyl fermions with opposite chiralities, and also opposite
eigenvalues under the $Z_2$ symmetry: \beqn H = \int d^3x \ v
\psi^\dagger_L \left( i \vec{\sigma}\cdot \vec{\partial}
\right)\psi_L - v \psi^\dagger_R \left( i \vec{\sigma}\cdot
\vec{\partial} \right)\psi_R. \label{edge3dx} \eeqn The charge
conservation $U(1)$ symmetry and the $Z_2$ symmetry act on the
boundary fermions as $ U(1): \psi_L, \psi_R \rightarrow
e^{i\theta}\psi_L, e^{i\theta} \psi_R$, $Z_2: \psi_L \rightarrow
\psi_L, \ \ \psi_R \rightarrow - \psi_R$. With this $U(1)\times
Z_2$ symmetry action, this TI has a $\mathbb{Z}$ classification
without interaction, because any fermion bilinear mass term should
either break the $Z_2$ symmetry (which corresponds to the ordinary
Dirac mass term that mixes the left and right handed Weyl
fermion), or break the U(1) symmetry, which is a fermion pairing
term (the so-called Majorana mass term).

To study the interaction effects on the TI, we follow the similar
procedure of Ref.~\onlinecite{senthilhe3}. We first couple the
left and right handed Weyl fermions both to a complex bosonic
order parameter $\phi$: \beqn \phi\left( \psi^\mathsf{T} \sigma^y \psi_L
+ \psi^\mathsf{T}_R \sigma^y \psi_R\right) + H.c. \label{phi}\eeqn When
$\phi$ condenses, $i.e.$ $\langle \phi \rangle \neq 0$, the charge
$U(1)$ symmetry is spontaneously broken and the fermions are all
gapped. Then we try to restore the $U(1)$ symmetry by
proliferating the vortex loops of the $U(1)$ order parameter
$\phi$. Along the vortex loop, by directly solving the Dirac
equation with a background vortex configuration of $\phi$, we will
find there are non-chiral $1d$ Majorana fermion described the
effective Hamiltonian (Fig.~\ref{fig}$b$) \beqn H_{1d} = \int dx \
i v (\chi_L
\partial_x \chi_L - \chi_R
\partial_x \chi_R). \label{H1d} \eeqn Again the left and right moving
Majorana fermions carry eigenvalue $\pm 1$ under the $Z_2$
symmetry respectively.

The $1d$ Hamiltonian Eq.~\ref{H1d} and its $Z_2$ symmetry are
exactly the same as the $1d$ edge states of $2d$ $p_x \pm ip_y$
TSC, where the $p_x + ip_y$ and $p_x - ip_y$ superconductors carry
eigenvalues $\pm 1$ respectively under the $Z_2$ symmetry. It is
known that for eight copies of Eq.~\ref{H1d}, a local interaction
can gap out Eq.~\ref{H1d} without breaking the $Z_2$
symmetry~\cite{qiz8,yaoz8,zhangz8,levinguz8} or ground state
degeneracy. Thus the $U(1)\times Z_2$ symmetry can be restored for
eight copies of the $3d$ edge states Eq.~\ref{edge3dx} by
proliferating the fully gapped vortex loops. Hence the
classification of the $D=4$ TI described above is reduced from
$\mathbb{Z}$ to $\mathbb{Z}_8$ under local interactions.

Now let's consider the case with $D = 4$, $d = 3$, namely three
out of the four dimensions are physical, while one dimension is
synthetic. We will still look at the most natural boundary of the
synthetic TI which is parallel to the synthetic dimensions. Then
for $N$ copies of the system, at each fixed synthetic momentum
$p$, the Hamiltonian for the boundary states reads \beqn H_p =
\int d^2x \ \sum_{a = 1}^N v \psi^\dagger_{a,L} \left( i \sigma^x
\partial_x + i \sigma^y
\partial_y  - p \sigma^z \right)\psi_{a,L} \cr\cr -
v \psi^\dagger_{a,R}\left( i \sigma^x \partial_x + i \sigma^y
\partial_y  - p \sigma^z \right)\psi_{a,R} + H_{int}(p). \label{edge4} \eeqn
The Hamiltonian Eq.~\ref{edge4} can be viewed as the previously
discussed two dimensional non-chiral TI with a $U(1)\times Z_2$
symmetry tuned close to the topological-trivial transition point
$p = 0$, and $p$ serves as the tuning parameter for this
transition. As we have already mentioned before, the
classification of this two dimensional non-chiral TI is reduced
from $\mathbb{Z}$ to $\mathbb{Z}_4$ under
interaction~\cite{fuz4,hermelez4}. Hence for $N=4$, the critical
point $p = 0$ can be avoided under interaction without breaking
the $Z_2$ symmetry, and the two regions with $p < 0$ and $p > 0$
can be smoothly connected without closing the gap. This implies
that the classification of the $D=4$ and $d=3$ synthetic TI is
reduced to $\mathbb{Z}_4$ rather than $\mathbb{Z}_8$ under the
generic interaction of the synthetic TI.

If we instead choose $D = 4$, $d = 2$ in the first place, namely
two out of the four dimensions are synthetic dimensions, then at
every synthetic momentum $\vec{p} = (p_x, p_y)$, the Hamiltonian
reads \beqn H_p &=& \int dx \ \psi^\dagger_{L} \left( i \sigma^x
\partial_x - p_y \sigma^y - p_z \sigma^z \right)\psi_{L} \cr\cr &-&
\psi^\dagger_{R} \left( i \sigma^x \partial_x - p_y \sigma^y - p_z
\sigma^z \right)\psi_{R}. \label{edge42} \eeqn Now at each $p_z$,
the edge state can be viewed as the $D=2$, $d=1$ synthetic TI with
$U(1)\times Z_2$ symmetry tuned close to the topological-trivial
transition, while one of the dimension is a synthetic dimension
$p_y$. $p_z$ is precisely the tuning parameter that tunes the two
dimensional synthetic system across the transition. As we have
argued before, the classification of this $D=2$ synthetic TI is
reduced to $\mathbb{Z}_2$ under generic synthetic interaction.
This implies that for two copies of Eq.~\ref{edge42}, a properly
designed $H_{int}(\vec{p})$ would gap out the edge system
Eq.~\ref{edge42} without ground state degeneracy. Then the
classification reduces from $\mathbb{Z}$ to $\mathbb{Z}_2$ when
$\delta = d = 2$.

To summarize this section, for a $D = 4$ non-chiral interacting
synthetic TI with $U(1)\times Z_2$ symmetry and different choices
of $\delta = D - d$, its classification is \beqn \mathbb{Z}_8, \
(\delta = 0); \ \  \ \mathbb{Z}_4, \ (\delta = 1); \ \ \
\mathbb{Z}_2, \ (\delta = 2). \eeqn

\section{Sythentic TI in general even dimensions \boldmath $D=2n$ }
\label{sec:D=2n}

In general even spatial dimensions $D=2n$, we can consider the
generalized version of the non-chiral TI with $U(1)\times Z_2$
symmetry discussed in Sec. \ref{sec:D=2} and \ref{sec:D=4}. Such
TIs in the non-interacting limit can always be constructed by
putting the fermions with $Z_2$ eigenvalue $+1$ into a
$2n-$dimensional Chern insulator with a non-trivial Chern-Simons
response, while simultaneously, putting fermions with $Z_2$
eigenvalue $-1$ also into a Chern insulator but with an opposite
Chern-Simons response. In the non-interacting limit, the
classification of this type of TI is given by $\mathbb{Z}$.

Let us first consider the case with all of the dimensions
physical, namely $D=d=2n$ and $\delta=0$. The boundary state of
this non-chiral TI
can be described by the Hamiltonian \beqn H = \int
d^{2n-1}x \sum_{k=1}^{2n-1} v\left( \psi^\dagger_L \left( i
\gamma^k \partial_k \right)\psi_L - \psi^\dagger_R \left( i
\gamma^k \partial_k \right) \psi_R \right). \label{edge2n-1dx}
\eeqn where $\{\gamma^k\}$ with $k=1,2,...,2n-1$ are
$2^{n-1}$-dimensional matrices that generate the complex Clifford
algebra. The products of all $\gamma$ matrices $\prod_{k=1}^{2n-1}
\gamma^k$ is a constant. The symmetries act on the boundary
fermions as $ U(1): \psi_L, \psi_R \rightarrow e^{i\theta}\psi_L,
e^{i\theta} \psi_R$, $Z_2: \psi_L \rightarrow \psi_L, \ \ \psi_R
\rightarrow - \psi_R$. The $U(1)\times Z_2$ symmetry forbids any
fermion bilinear mass terms because all ordinary mass terms break
the $Z_2$ symmetry due to the mixing of $\psi_L$ and $\psi_R$
fermions and all possible Majorana mass terms break the $U(1)$
symmetry. Therefore, this TI has a $\mathbb{Z}$ classification in
the non-interacting limit.

In the following, we will show that the classification of
$U(1)\times Z_2$ non-chiral TIs in $D=2n$ spatial dimensions with
$\delta=0$ reduces from $\mathbb{Z}$ to $\mathbb{Z}_{2^{n+1}}$
when we consider interactions. To show this, we apply the method
of fermion sigma model\cite{youxu2}. We first consider $\nu=2^n$
copies of the boundary states described in Eq. \ref{edge2n-1dx},
rewrite them using Majorana fermions $\chi$ and couple the
Majorana fermions to a $(2n+2)$-component dynamical vector field
$n_a$ ($a=1,...,2n+2$) via the mass terms: \beqn H_{\times \nu} =
\int d^{2n-1}x~  \chi^{\sfT} \left(  \sum_{k=1}^{2n-1} i v
\alpha^k \partial_k + \sum_{a=1}^{2n+2} n_a \beta^a
 \right) \chi. \label{edge2n-1dx_nu}
\eeqn Here, the set of matrices $\{\alpha^1,..,\alpha^{2n-1};
i\beta^1,...,i\beta^{2n+2}\}$ generates the real Clifford algebra
$\cl_{2n-1,2n+2}$ in its $2^{2n+1}$-dimensional representation.
The matrices $\alpha^i$ are real symmetric matrices while the
matrices $\beta^a$ are imaginary anti-symmetric ones. All the
$\alpha^i$ and $\beta^a$ matrices square to the identity matrix.
In this representation, the $U(1)$ and $Z_2$ symmetry actions on
the Majorana fermions $\chi$ are generated by $i\beta^1\beta^2$
and $\prod_{i=1}^{2n-1}\alpha^i$ respectively. The symmetry
transformation of the vector field $n_a$ is given by \beqn
 U(1): &
(n_1+i n_2) \rightarrow e^{i\theta} (n_1+i n_2);  \nonumber \\
&  n_a \rightarrow n_a~~\text{for}~ a=3,...,2n+2.
\label{Eq:sym_action_n_nc}
 \\
 Z_2:&  n_a \rightarrow -n_a ~~\text{for all}~ a. \nonumber
\eeqn Assuming that the vector field $n_a$ is a slowly varying
field with unit modulus, we can integrate out the Majorana
fermions $\chi$ and obtain an effective action of $n_a$ which can
be identified as a $O(2n+2)$ non-linear sigma model ($\NLSM$) in
$2n-1$ spatial dimensions with a level-$1$ (or level-$-1$)
Wess-Zumino-Witten (WZW) term\cite{youxu2,abanov}. Such an
effective action can be identified as that of the boundary state
of a bosonic symmetry protected topological (SPT) state with $U(1)
\times Z_2$ symmetry. The bulk of such a bosonic SPT should be
described by a $O(2n+2)$ $\NLSM$ in $2n$ spatial dimensions with a
$2\pi$ $\Theta-$term. Now, we can conclude that $\nu$ copies of
the TI whose boundary state is described in Eq. \ref{edge2n-1dx}
can be adiabatically connected to this bosonic SPT state. This
bosonic SPT has a $\mathbb{Z}_2$ classification because we can
consider two copies of the $O(2n+2)$ $\NLSM$s each with a $2\pi$
$\Theta$-term and couple them such that their $n_a$ vector fields
aligns in the first $2n+1$ components and anti-aligns in the
$2n+2$th component. The coupling effectively merges the two
copies into a single $O(2n+2)$ $\NLSM$s which is free of a net
$\Theta$-term and hence is topologically trivial. Therefore, we
can conclude that $2\nu$ copies of the boundary state described in
Eq. \ref{edge2n-1dx} together are topologically equivalent to the
boundary state of a trivial state. The classification of the $U(1)
\times Z_2$ non-chiral TI is hence reduced from $\mathbb{Z}$ in
the non-interacting limit to $\mathbb{Z}_{2\nu} =
\mathbb{Z}_{2^{n+1}}$ in $D=2n$ spatial dimensions (with
$\delta=0$) when we take interactions into account.

Now, we consider the case with one of the spatial dimension
synthetic, i.e. $D=2n$, $d=2n-1$ and $\delta=1$. We can start with
the boundary state Hamiltonian: \beqn H_p = &  \int d^{2n-2}x
\left\{ \ \psi^\dagger_L \left( \sum_{k=1}^{2n-2} i v\gamma^k
\partial_k - p \gamma^{2n-1} \right)\psi_L \right.
\nonumber \\
& - \left. \psi^\dagger_R \left( \sum_{k=1}^{2n-2} i v\gamma^k
\partial_k - p \gamma^{2n-1} \right)\psi_R \right\}
\label{edge2n-1dx_delta1}, \eeqn where $p$ represents the
synthetic momentum along the synthetic dimension. We notice that
this boundary Hamiltonian Eq. \ref{edge2n-1dx_delta1} can be
identified as the bulk Hamiltonian of the $U(1)\times Z_2$
symmetric non-chiral TI in $D'=d'=2n-2$ spatial dimensions tuned
to the vicinity of a topological-trivial transition point at
$p=0$. As we just discussed, the classification of the
$D'=d'=2n-2$ non-chiral TI is reduced from $\mathbb{Z}$ to
$\mathbb{Z}_{2^n}$ under interaction. That means when we consider
$2^{n}$ copies of the model Eq. \ref{edge2n-1dx_delta1}, the
critical point at $p=0$ can be avoided under interactions without
breaking the $U(1)\times Z_2$ symmetry. In another word, the
classification of the $U(1)\times Z_2$ symmetric non-chiral TI in
$D=2n$ dimensions with $\delta=1$ is reduced to
$\mathbb{Z}_{2^n}$.

For $\delta \geq 1$, we can always perform a similar analysis by
studying the boundary state Hamiltonians that are counterparts of
Eq. \ref{edge2n-1dx} and Eq. \ref{edge2n-1dx_delta1}.
Interestingly, such boundary state Hamiltonians can always be
identified as the bulk Hamiltonians of the $U(1)\times Z_2$
symmetric non-chiral TI in $D'=2n-2$ dimensions with
$\delta'=\delta-1$ near the topological-trivial critical point.
This identification is done by choosing one of the synthetic
momenta as the tuning parameter for the topological-trivial
transition while leaving the other $\delta'=\delta-1$ synthetic
momenta still representing the synthetic dimensions among the
total $D'=2n-2$ dimensions. With this identification, we can
directly conclude that, for general $\delta$, the classification
of the $U(1)\times Z_2$ synthetic non-chiral TI becomes $\mathbb{Z}_{2^{n+1-\delta}}$.

To summarize this section, for a $2n$-dimensional non-chiral
interacting synthetic TI with $U(1)\times Z_2$ symmetry, the
classification, in the presence of $\delta$ synthetic dimensions,
is given by $\mathbb{Z}_{2^{n+1-\delta}}$. This result is
consistent with the analysis in Sec. \ref{sec:D=2} and
\ref{sec:D=4}.

\section{Sythentic TI in general odd dimensions \boldmath $D=2n+1$ }
\label{sec:D=2n+1} In general odd spatial dimensions $D=2n+1$, we
can consider the generalized version of the TI with $U(1)\times
Z_2^T$ symmetry discussed in Sec. \ref{sec:D=3}. In the
non-interacting limit, such systems belong to class AIII in the
``10-fold way" classification\cite{ludwigclass1,ludwigclass2} and
have a classification of $\mathbb{Z}$.

Let's first consider the case with all of the dimensions physical,
namely $D=d=2n+1$ and $\delta=0$. The boundary state of the
$U(1)\times Z_2^T$ symmetric TI that can generate the whole
$\mathbb{Z}$ class can be described by the Hamiltonian \beqn H =
\int d^{2n}x \sum_{k=1}^{2n} v \psi^\dagger \left( i \gamma^k
\partial_k \right)\psi, \label{edge2ndx} \eeqn where
$\{\gamma^k\}_{k=1}^{2n}$ are a set of $2^{n}$-dimensional complex
matrices that forms the complex Clifford algebra. The symmetries
act on the boundary fermions as $ U(1): \psi \rightarrow
e^{i\theta}\psi$, $Z_2^T: \psi \rightarrow \left(\prod_{i=1}^{2n}
\gamma^i \right) \psi^\dag$. One can check that the $U(1)\times
Z_2^T$ symmetry forbids any fermion bilinear mass terms even when
there are multiple copies of such boundary states. Therefore, this
TI has a $\mathbb{Z}$ classification in the non-interacting limit.

In the following, we will show that the classification of
$U(1)\times Z_2^T$ TI in $D=2n+1$ spatial dimensions with
$\delta=0$ reduces from $\mathbb{Z}$ to $\mathbb{Z}_{2^{n+2}}$
when we take interactions into account. To show this, we will
again use the method of fermion sigma model. The following
discussion will be largely parallel to Sec. \ref{sec:D=2n}. We
first consider $\nu'=2^{n+1}$ copies of the boundary states
described in Eq. \ref{edge2ndx}, rewrite them using Majorana
fermions $\chi$ and couple the Majorana fermions to a
$(2n+3)$-component dynamical vector field $n_a$ via the mass terms
\beqn H_{\times \nu'} = \int d^{2n}x~  \chi^{\sfT} \left(
\sum_{k=1}^{2n} i v \alpha^k \partial_k + \sum_{a=1}^{2n+3} n_a
\beta^a
 \right) \chi. \label{edge2ndx_nu'}
\eeqn Here, the set of matrices $\{\alpha^1,..,\alpha^{2n};
i\beta^1,...,i\beta^{2n+3}\}$ generates the real Clifford algebra
$\cl_{2n,2n+3}$ in its $2^{2n+2}$-dimensional representation. In
this representation, the $U(1)$ symmetry action on the Majonara
fermions is generated by $i\beta^1\beta^2$. The anti-unitary
time-reversal symmetry acts as $Z_2^T: \chi \rightarrow \left(
\prod_{i=1}^{2n } \alpha^i \right) \chi$, combined with a complex conjugation. The symmetry
transformation of the vector field $n_a$ is given by \beqn
 U(1): &
(n_1+i n_2) \rightarrow e^{i\theta} (n_1+i n_2);  \nonumber \\
&  n_a \rightarrow n_a~~\text{for}~ a=3,...,2n+3.
\label{Eq:sym_action_n_ac}
 \\
 Z_2^T:&  n_a \rightarrow -n_a ~~\text{for all}~ a. \nonumber
\eeqn As we discussed in Sec. \ref{sec:D=2n}, when the vector
field $n_a$ is slowly varying and has a unit modulus, we can
integrate out the Majorana fermion $\chi$ and obtain a $O(2n+3)$
$\NLSM$ with a level-1 (or level-$-1$) WZW term. Here, the action
of $n_a \rightarrow -n_a$ in $O(2n+3)$ should be identified as the
time-reversal symmetry $Z_2^T$. Such an effective action in fact
describes the boundary state of a $U(1)\times Z_2^T$ symmetric
bosonic SPT which itself has a $\mathbb{Z}_2$ classification.
Therefore, by the same argument given in Sec. \ref{sec:D=2n}, the
classification of the $U(1) \times Z_2^T$ symmetric TI is reduced
from $\mathbb{Z}$ in the non-interacting limit to
$\mathbb{Z}_{2\nu'} = \mathbb{Z}_{2^{n+2}}$ in $D=2n+1$ spatial
dimensions (with $\delta=0$) when we take interactions into
account.

When synthetic dimensions are present, i.e. $\delta \geq 1$, we
can identify one of the synthetic momenta as the tuning parameter
of a topological-trivial transition, the boundary state
Hamiltonian can then be identified as the bulk Hamiltonian of a
$U(1) \times Z_2^T$ TI in the vicinity of the topological-trivial
critical point in $D'=2n-1$ spatial dimensions and with
$\delta'=\delta-1$ synthetic dimensions. With this identification,
we can conclude that the classification of $U(1) \times Z_2^T$
symmetric interacting TI is given $Z_{2^{n+2-\delta}}$.

To summarize this section, for a $2n+1$-dimensional interacting
synthetic TI with $U(1)\times Z_2^T$ symmetry, the classification,
in the presence of $\delta$ synthetic dimensions, is given by
$\mathbb{Z}_{2^{n+2-\delta}}$. This result is consistent with the
analysis in Sec. \ref{sec:D=3}.



\section{Summary}


In this work we have analyzed the interaction effects on the
effective TIs simulated with the newly developed synthetic
techniques. We demonstrate that unlike ordinary interacting TIs,
the interaction causes different classification reduction of the
simulated TI, due to the generic while special form of the
interaction in systems with synthetic dimensions. We need to point
out that the analysis used the fact that the system at every
synthetic momentum $\vec{p}$ is a lower dimensional system with
the same symmetry as the desired effective $D-$dimensional TI.
This analysis no longer naturally applies (although not
impossible) if the system involves time-reversal symmetry that
does not commute with the charge $U(1)$ symmetry, because in this
case time-reversal symmetry would bring $c_k$ to $c_{-k}$, hence
it mixes systems labelled by different parameter $\vec{p}$.

Topological semimetals have also attracted enormous research
interests and efforts in the last few
years~\cite{weylashvin,weylleon,weyldai,weylreview,weylex1,weylex2,diracex1}.
One can also construct semimetals in lab using the same synthetic
techniques, and these semimetals could be vulnerable to
interactions, $i.e.$ the bulk of the system can be driven into an
insulator with nondegenerate ground state due to interaction. For
example, the long wave-length effective Hamiltonians
Eq.~\ref{edge2dx} and Eq.~\ref{edge4} can also be viewed as the
bulk Hamiltonian expanded near the gapless momenta of synthetic
semimetals with $D=2$ and $D=3$, again $p$ is the tuning parameter
which is viewed as an extra synthetic momentum. And our previous
analysis indicates that for $N=4$, both cases can be gapped out by
interaction $H_{int}(p)$ without leading to ground state
degeneracy, and the Dirac semimetal becomes an insulator with
nondegenerate ground state.

It has also been shown that under strong interaction the boundary
of a nontrivial $3d$ TI can be a $2d$ topological order with
anomalous quantum number
fractionalizations~\cite{TOQi,TOSenthil,TOAshvin,TOMax,TOTeo,TOYe,TOLevin,TOCheng1,TOCheng2}.
As we explained before, at each parameter $p$, the edge
Hamiltonian Eq.~\ref{edge4} describes an interacting $2d$ system
at the boundary of a $D=4$ system. Thus under strong interaction
it is possible that at each parameter (effective momentum) $p$ the
edge is driven into a topological order. This situation could
correspond to a very exotic ``topological order" at the boundary
of the $D = 4$ TI. We will leave this topic for future studies.

The authors are supported by the David and Lucile Packard
Foundation and NSF Grant No. DMR-1151208.

\bibliography{syn}

\end{document}